\documentstyle[prl,aps,multicol]{revtex}
\begin{document}
\draft
\title{Field theory of the quantum kicked rotor}
\author{Alexander Altland\cite{perm} and Martin R. Zirnbauer\cite{perm}}
\address{Institute for Theoretical Physics, University of California, 
Santa Barbara, U.S.A.}
\maketitle
\begin{abstract}
  The quantum kicked rotor (QKR) is investigated by field
  theoretical methods. It is shown that the effective theory
  describing the long wave length physics of the system is precisely
  the supersymmetric nonlinear $\sigma$--model for quasi
  one--dimensional metallic wires. This proves that the analogy
  between chaotic systems with dynamical localization and disordered 
  metals can indeed be exact. The role of symmetries is discussed.
\end{abstract}
\begin{multicols}{2}
\narrowtext Quantum mechanics tends to suppress the chaoticity of
classical dynamical systems. The investigation of this phenomenon in
{\it periodically driven} systems, i.e. systems that are governed by a
Hamiltonian with periodic time dependence, has led to the discovery of
one of the most intriguing parallels between the fields of nonlinear
dynamics and disordered solids: In the quantum kicked rotor (QKR), a
typical representative of this class of system, the quantum mechanical
suppression of chaos exhibits striking similarities to the phenomenon
of Anderson localization in disordered metallic wires\cite{ccfi}. \par
By definition, the kicked rotor is a point particle that moves freely
on a circle. The particle is kicked periodically in time, where the
kick strength depends on the angular position. When a kick strength
parameter $k\tau$ exceeds a certain threshold value, the dynamics
becomes globally chaotic. In a statistical physicist's language, the
chaoticity of the motion manifests itself as follows: An ensemble of
particles prepared at time $t=0$ so as to have definite angular
momentum $l_0$ but arbitrary angular coordinate $\theta$, will {\it
  diffuse} in $l$--space around the initial condition $l_0$.  In the
corresponding {\it quantum system} ($\theta \rightarrow \hat{\theta},
\; l \rightarrow \hat{l},\; [\hat{l}, \hat{\theta}] =-i \hbar$), the
unbound diffusion in $l$--space is suppressed by localization.
Numerical\cite{ccfi} and analytical\cite{fgp,aa} studies have shown
that the QKR--localization is analogous to the Anderson localization
displayed by metallic wires with many channels (quasi $1d$ wires).  In
particular, it has been demonstrated numerically\cite{cim} that a
phenomenological modeling of the QKR by random band matrices can
successfully explain essential features of the localization
phenomenon. Random matrix models of the same type are known\cite{fm},
in turn, to describe the universal large distance physics of
disordered wires.
\par
However, the equivalence between the rotor and quasi $1d$ wires 
still has the status of a conjecture. A rigorous answer to the
question whether this analogy is {\it complete} (and not just
restricted to the bilateral appearance of localization) has not
yet been given. Can a simple one--dimensional driven system indeed
exactly mimic the behavior of disordered electronic conductors,
which includes a variety of complex phenomena that have recently been
found\cite{fe,fm2,km}? In the present Letter we are going to show that
the answer is positive.  This is done by mapping the kicked rotor onto
the very same supersymmetric nonlinear $\sigma$--model that is known
to describe the long wave length physics of disordered wires. The fact
that both models can be described by the same effective field theory
implies that all that is known about the QKR applies to disordered
wires and vice versa.\par
The QKR is defined by the time dependent Hamiltonian
\[
\hat{H}=\frac{\hat{l}^2}{2}+k\cos(\hat{\theta}+a)
\sum_{n=-\infty}^{\infty}\delta(n\tau-t),
\]
where the particle's moment of inertia has been set to unity and $a\in
{\cal R}$ is a symmetry breaking parameter whose meaning will be
explained below. To elucidate the analogy between this Hamiltonian and
disordered electron systems, one may consider the discrete time analog
of a four--point Green function in (angular) momentum space:
\begin{equation}
\Big\langle
\left\langle l_1|G^+(\omega_+)|l_2\right\rangle
\left\langle l_3|G^-(\omega_-)|l_4\right\rangle
\Big\rangle_{\omega_0},
\label{K}
\end{equation}
where $G^{\pm}(\omega_{\pm}):= \sum_{n=0}^{\pm\infty} \hat{U}^n e^{i
  \omega_{\pm} n\tau}= [1-(\hat{U} e^{i \omega_{\pm} \tau} )^{\pm1}
]^{-1}$, $\hat{U}= \exp(i\hat{l}^2\tau/4) \exp(ik \cos(\hat{\theta}+a))
\exp(i\hat{l}^2\tau/4)$ denotes the Floquet operator, i.e. the unitary
operator governing the time evolution during one elementary time step,
$\omega_{\pm}=\omega_0\pm(\omega/2+i0)$, and $\langle\dots
\rangle_{\omega_0} := \tau \int_0^{2\pi/\tau} d\omega_0(\dots) /2\pi$ 
is an average over the rotor's quasi--energy spectrum.\par
Before turning to the quantitative discussion of the system, let us
explain the meaning of the symmetry breaking parameter $a$. The
localization length of a disordered wire depends on the behavior of
the Hamiltonian under the time reversal transformation $T$: $t\rightarrow
-t$, $p \rightarrow -p$, $x\rightarrow x$, where $p$ and $x$ are the
momentum and the position. In the case of the rotor, where
localization takes place in momentum rather than coordinate space,
this symmetry operation is irrelevant. However, it has been
shown\cite{bs} that the transformation $T_c$: $t \rightarrow -t$,
$\theta \rightarrow -\theta$, $l \rightarrow l$ plays a role analogous
to $T$ in disordered metals. (Note that $T_c$ differs from $T$ just by
the exchange of momentum and position.) To couple the system to a
$T_c$ breaking perturbation in a simple way, we put it on an angular
lattice of spacing $1/(2\pi L)$, $L\in {\cal N}$, thereby giving it 
the topology of a ring of circumference $L$ in {\it momentum
space}\cite{i}. It will turn out that the symmetry breaking
parameter $a$ then acts like a $T_c$ breaking Aharonov--Bohm flux
piercing the ring.\par
In the following we deal with the correlator (\ref{K}) by field
theoretical methods. The strategy of our approach is dictated by the
experience gained from both the analysis of disordered metals\cite{e}
and a recent field theoretical approach\cite{aaas} to Hamiltonian chaotic
systems. Owing to the different formulation of periodically
driven systems, however, the actual computational scheme deviates
significantly from these cases. To simplify the notation, we
temporarily focus on the case of unbroken $T_c$ symmetry, $a=0$.\par
Invariance under the transformation $T_c$, which acts as an {\it
anti--unitary} operator in the quantum system, results in the Floquet 
operator being a symmetric matrix when represented in the $l$--basis 
(from now on we refer to all operators in $l$--representation). This 
makes it possible to decompose $U=\{\langle l|\hat{U} |l' \rangle \}$ 
by $( U e^{i\omega_\pm \tau})^{\pm 1} = V_\pm^{\vphantom{T}} V_\pm^T$, 
where $V_\pm$ does not possess any symmetries other than unitarity. We 
choose $V_\pm = e^{\pm i\omega_\pm \tau/2} K^{a=0}_{\pm,k/2}$, where 
$K^{a}_{\pm,k} := 
\{\langle l| \exp(\pm i\hat{l}^2\tau/4) \exp(\pm ik \cos(\hat{\theta}+a))|l'
\rangle \}$, and write the Green functions appearing in (\ref{K}) as\cite{fn1}
\begin{equation}
G^{\pm}(\omega_\pm) = \left(\begin{array}{cc}
1  &V_\pm\\
V_\pm^T&1
\end{array}\right)^{-1}_{11}=:\tilde{G}^\pm(\omega_\pm)_{11} .
\label{Gdoubled}
\end{equation}
In the next step we
introduce a superfield $\psi=\{\psi_{\lambda \alpha t l} \}$,
$\lambda, \alpha, t = 1, 2$, with complex commuting (anticommuting)
components $\psi_{\alpha=1}(\psi_{\alpha=2})$, and consider the
generating functional
\begin{equation}
\int {\cal D}(\psi,\bar\psi)
\exp\left[-\bar\psi\left({G}^{-1}+{J}\right)\psi\right] ,
\label{funint1}
\end{equation}
where ${G}= E_{\rm AR}^{11} \otimes 1_{\rm BF}^{\vphantom{1}} \otimes
\tilde{G}^+(\omega_+) + E_{\rm AR}^{22} \otimes 1_{\rm BF}^{\vphantom{1}}
\otimes\tilde{G}^-(\omega_-)$, matrices with subscript 'AR' ('BF','T')
act in the two--dimensional spaces of $\lambda$ ($\alpha$,$t$) indices
(the $t$--indices refer to the matrix structure appearing in
(\ref{Gdoubled})) and $(E_{\rm X}^{ij})_{i'j'} := \delta_{ii'}
\delta_{jj'},\; {\rm X=AR,BF,T}$. Here and below, indices that are
not indicated explicitly are summed over. Expressions like (\ref{K})
can readily be obtained from (\ref{funint1}) by differentiating twice
with respect to matrix elements of the source field ${J}$. As we
are interested in the general structure of the theory, rather than in
the calculation of any particular correlation function, we henceforth
omit ${J}$.\par
After a few elementary manipulations, namely matrix transpositions and
regrouping of integration variables, the Gaussian integral (\ref{funint1})
takes the simple form
\begin{equation}
\label{funint2}
\int {\cal D}(\phi,\chi)
e^{-\frac{1}{2}(\bar\phi\phi+\bar\chi\chi)+
\bar\phi E^{11}_{\rm AR} \otimes V_+ \chi +
\bar\chi E^{22}_{\rm AR} \otimes V_-^T \phi },
\end{equation}
where the fields $\phi=\{\phi_{\lambda \alpha t l}\}$ and
$\chi=\{\chi_{\lambda \alpha t l}\}$ comprise components of both
$\psi$ and $\bar\psi$. Instead of displaying the structure of these
new quantities explicitly, we merely note two essential
features that fix their functionality as integration variables: (i)
$\phi$ and $\chi$ are independent of each other and (ii) they possess
the symmetry ${\bar Y} = (1_{\rm AR}\otimes M)Y,\; Y=\phi, \chi$,
where $M=E^{11}_{\rm BF}\otimes \sigma^1_{\rm T}+ E^{22}_{\rm BF}
\otimes (i\sigma^2_{\rm T})$ and $\sigma_{\rm X}^i$ $(i=1,2,3,\;
{\rm X=AR,BF,T})$ denotes the Pauli matrices.\par
The next step in the construction of the field theory is
the average over the phase $\exp(i\omega_0\tau)$, which plays a role
similar to the energy average employed in Ref.\cite{aaas}. In that
case, energy averaging led to a quartic ($\sim(\bar\psi\psi)^2$)
non--local contribution to the action of the field theory. The latter
was eliminated by means of a matrix valued auxiliary field $Q$ that
coupled to the {\it dyadic} product $\psi\bar\psi$.  
The phase average to be carried out in the present problem
produces in addition to the quartic term an infinite series of higher
contributions to the action. We have not succeeded in decoupling these
terms by elementary means. On the other hand, the experience gained
from previous diagrammatic analyses\cite{aa} of the QKR suggests that
a field coupling to $\psi\bar\psi$ should again describe the large
scale physics.\par 
The problem of identifying this field is solved by a recently
discovered identity\cite{z} that adapts the Hubbard--Stratonovich 
transformation to averages over {\it unitary} operators. In the 
special case under consideration, namely a phase or ${\rm U}(1)$ 
average, this identity reads:
\begin{equation}
\label{magic}
\left\langle e^{
\bar\phi_1^{\vphantom{\dagger}}
u \eta^{\vphantom{\dagger}}_1+
\bar\eta_2^{\vphantom{\dagger}}
{\bar u} \phi^{\vphantom{\dagger}}_2
}
\right\rangle_{\omega_0}=\int {\cal D}\mu(Z,\tilde{Z})
e^{\bar\phi_1^{\vphantom{\dagger}} Z \phi_2^{\vphantom{\dagger}}+
   \bar\eta_2^{\vphantom{\dagger}} \tilde{Z}\eta_1^{\vphantom{\dagger}}},
\end{equation}
where $u := \exp(i\omega_0\tau)$, $\eta_1 = V_+
\big|_{\omega_0=0} \chi_1$, $\bar\eta_2 = \bar\chi_2
V_-^T \big|_{\omega_0=0}$, all subscripts refer to the
$\lambda$--indices ('AR'--space), $Z = \{ Z_{\alpha t l,\alpha' t' l'} \}$
is a non--local (in $l$) $4\times4$ supermatrix field, ${\cal D}
\mu(Z,\tilde{Z})={\cal D}(Z,\tilde{Z}) {\rm
sdet}(1-Z\tilde{Z}$) with 'sdet' the
superdeterminant, and $\int {\cal D}(Z,\tilde{Z})$ stands for the
integral over the matrix elements of $Z$ and $\tilde{Z} :=
Z^{\dagger}\sigma^3_{\rm BF}$.\par
The proof\cite{z} of (\ref{magic}) makes use of group theoretical
concepts and the theory of generalized coherent states\cite{p} and is
too lengthy to be reported here. We note however that the field
$Z_{ll'}$ takes values in the unrestricted set of $4\times4$ complex
supermatrices. The latter can be interpreted as a space parameterizing
the coset space ${\bf G}/{\bf K}, \;{\bf K}=\{k\in{\bf G}|
k\sigma^3_{\rm AR}=\sigma^3_{\rm AR}k\}\subset {\bf G}$, where $\bf G$
is the group of $8\times8$ supermatrices $g$ subject to the constraint
$g^{\dagger}\eta g =\eta$, $\eta = \left( \sigma^3_{\rm AR} \otimes
E^{11}_{\rm BF} + 1_{\rm AR}^{\vphantom{1}} \otimes E^{22}_{\rm
BF}\right) \otimes 1_{\rm T}$. This coset space is the field manifold
of a '{\it unitary}' supersymmetric $\sigma$--model that is twice as
large as in the usual case\cite{e} on account of the extra T--space
indices. The relationship between the $Z$--field and this manifold is
the first indication of the fact that we will end up with a nonlinear
$\sigma$--model.\par
After this comment on the formalism, we 
proceed to apply (\ref{magic}) to the
construction of the field theory for the rotor. To that end we insert
(\ref{magic}) into (\ref{funint2}) and perform the Gaussian
integration over the fields $\phi$ and $\chi$. As a result we obtain
for the generating functional (at $J = 0$),
\begin{eqnarray}
&&\int{\cal D}(Z,\tilde{Z})\exp {\rm str}\bigg[{\rm ln}(1-Z\tilde{Z})
-\textstyle{1\over 2}{\rm ln}(1-Z\tau^{-1}Z^T\tau) \nonumber\\
&&\hspace{2.8cm}
-\textstyle{1\over 2}{\rm ln}(1-e^{i\omega\tau}\tilde{Z}U_0
\tau^{-1}\tilde{Z}^T\tau U_0^{\dagger})\bigg] ,
\label{Zfun1}
\end{eqnarray}
where $U_0=U\big|_{a=0}$, $\tau=M \sigma^3_{\rm BF}$ and the supertrace
'str' includes a trace over the $l$--space. So far all
manipulations have been exact. We next restrict the field theory to
its infrared limit, which describes the long time/large distance
physics we are interested in. \par
The action of the field theory (\ref{Zfun1}) vanishes for fields $Z$
proportional to unity in angular momentum space if $\omega\rightarrow 0$ 
and the constraint
\begin{equation}
\tilde{Z}=\tau^{-1}Z^{T}\tau
\label{Zsymmetry}
\end{equation}
is imposed.
Field configurations that violate this symmetry are 'massive' and
cannot contribute to the long range correlations of the model. We
therefore restrict the integration in (\ref{Zfun1}) to the field
manifold specified by (\ref{Zsymmetry}). (The integration over the massive
quadratic fluctuations around this manifold yields a factor of unity
by the supersymmetry of the model.) Note that the unitary coset space
${\bf G}/{\bf K}$ subject to the constraint (\ref{Zsymmetry}) defines
the field space of the '{\it orthogonal}' nonlinear $\sigma$--model.
As a preliminary result, we thus find that our field theory has the
same symmetries as the one describing time reversal invariant
disordered metals. What happens when this symmetry is gradually broken
by the introduction of a small finite value of $a$?  In that case, the
decomposition of the time evolution operator has to be generalized to
$(U e^{i\omega_\pm \tau})^{\pm 1} = V_\pm^a V_\pm^{-aT}$, where $V_\pm^a 
= e^{\pm i\omega_{\pm}\tau/2 } K^a_{\pm,k/2}$.  All further steps can 
be repeated in essentially the same way as before and we again arrive at
(\ref{Zfun1}). The only change is that the fields appearing in the
action have undergone a 'gauge transformation' $Z_{ll'}\rightarrow
Z^a_{ll'}:=\exp(-ial\sigma^3_{\rm T}) Z_{ll'} \exp(ial'\sigma^3_{\rm
T})$, and similarly for $\tilde{Z}$.\par
To carry the analogy to disordered metals further, we need to expand
the action around the limit $Z_{ll'}=\delta_{ll'}Z_0$, $\omega=0$. 
Details of this somewhat tedious calculation will be presented
elsewhere\cite{aaz}. Here we restrict ourselves to a rough sketch of
the main ideas. We first subject the action in~(\ref{Zfun1}) to a
'semiclassical approximation'. The expansion parameter of this
approximation is $\hbar/\delta l\ll 1$, where $\delta l$ is the
typical angular momentum scale over which the relevant $Z$--fields
fluctuate\cite{fn2}. As a result, (i) the symbol 'str' in (\ref{Zfun1})
no longer includes a trace over angular momentum space but rather an
integral over the phase space coordinates $(l,\theta)$ of the {\it
  classical} rotor, (ii) $Z(l,l')$ is replaced by its Wigner transform
$Z(l,\theta)$ (we temporarily suppress the superscript $a$ in $Z^a$),
and (iii) $U_0^{\vphantom{\dagger}} Z U_0^{\dagger}$ is replaced by 
$Z_u$, where $f_u(l,\theta) = f(\theta+\tau(l+k\sin\theta), l +
 k\sin\theta)$ denotes the classical one time step evolution 
(standard map) of a phase space function.\par
The angular variable $\theta$ of the standard map is a
rapidly relaxing degree of freedom, which leads us to expect that only
$\theta$--independent field configurations contribute to the long
wave limit of the model. To formulate this statement in a
quantitative manner we do a Fourier transform, $Z(l,\theta) =
\sum_{m = -\infty}^{\infty}Z_m(l)\exp(im\theta)$, and observe that the 
non--zero modes $Z_{m\not=0}$ are 'massive'. Integration over these fields 
in Gaussian approximation yields an effective field theory for the
massless zero mode $Z(l):=Z_0(l)$.\par
We finally expand the action $S[Z]$ in terms of slowly fluctuating
fields.  This program is carried out most economically
in the $l$--Fourier space, i.e. in angular coordinates\cite{fn3}.  The
small parameters of this expansion scheme are $\omega\tau$ and the
characteristic 'momentum' $\phi$ ($\phi$ is an angular variable) of
slowly fluctuating fields $Z(\phi)$, which is of order $\phi\ll
k^{-1}\ll 1$.  Concerning the former, we note that the mean quasi--energy
level spacing of the model is $\Delta=2\pi(L\tau)^{-1}$. Since we are
interested in small frequencies of ${\cal O}(\Delta)$, we have 
$\omega\tau\sim L^{-1}\ll1$. To leading order in $\omega\tau$ and $\phi$ 
the action of the generating functional reads
\begin{eqnarray}
&&S[Z^a,\tilde{Z}^a]\simeq \int dx\Big[
- \frac{i\omega\tau}{2}{\rm str}(1-Z^a(x)\tilde{Z}^a(x))^{-1}-\nonumber\\
&&\hspace{2cm} \frac{D}{4}\partial_{x'}
\partial_x {\rm str\,ln}(1-Z^a(x)\tilde{Z}^a(x'))\big|_{x'=x}\Big] ,
\end{eqnarray}
where we have Fourier--transformed back to angular momentum space,
taken a continuum limit (i.e. the variable $x$ is a smoothed version
of the $l$--index, $\sum_l\rightarrow\int dx$) and $D=k^2/2+...$ is the
classical diffusion coefficient of the rotor\cite{cc}. (The dots
indicate oscillatory corrections\cite{rrw} to $D$ that result from 
elimination of the non--zero modes and are smaller than the leading 
term by powers of $k$.) Introducing an $8\times8$ matrix field $Q$ by
\[
Q=
\left(\begin{array}{cc}
1        &Z\\
\tilde{Z}&1
\end{array}\right)
\left(\begin{array}{cc}
1&0\\
0&-1
\end{array}\right)
\left(\begin{array}{cc}
1        &Z\\
\tilde{Z}&1
\end{array}\right)^{-1} ,
\]
we can rewrite the functional integral as
\begin{eqnarray}
&&\int {\cal D}Q\exp \int {\rm str}\left(
\frac{D}{32}\nabla_a Q\nabla_a Q
+ \frac{i\omega\tau}{8}  Q\sigma^3_{\rm AR}\right) ,
\nonumber\\
&&\hspace{1.0cm}\nabla_a=\nabla +i a[\sigma^3_{\rm T},\;.\;],
\label{finalaction}
\end{eqnarray}
which is precisely the nonlinear $\sigma$--model for a
quasi one--dimensional metallic ring in the presence of a
$T$--breaking (Aharonov-Bohm) vector potential of strength $a$.\par
Because of its importance for an understanding of the localization
physics of wires, the model (\ref{finalaction}) has been investigated
thoroughly\cite{e}. Let us now review some of its essential
properties in the terminology of the kicked rotor.  For times less than
${\cal O}(\tau k^2)$ the kicked particle performs a diffusive motion in
momentum space. On larger time scales, quantum localization confines
the particle to stay within a volume specified by the localization
length $\xi_o=k^2/2$.  For 'flux' strengths $a\sim L^{-1}$, the
orthogonal symmetry of the model is broken and one expects a doubling
of the localization length $\xi_o\rightarrow
\xi_u=2\xi_o$\cite{e}. (Note, however, that $a_{\rm max}=2\pi/L$
corresponds to one 'flux quantum' $\phi_0$ penetrating the ring. As
the physics of Aharonov-Bohm geometries is $\phi_0$--periodic,
$a_{\rm max}$ is the maximum field strength that can be realized in our
model. For systems with $L\gtrsim\xi_o$ this strength does not suffice
to cause a crossover from $\xi_o$ to $\xi_u$.).\par
A careful look reveals that the localization length $\xi_o$ predicted
by our analysis is four times larger than the length $\xi_n$ found in
numerical work (cf. e.g. Ref.~\cite{s}). We believe that this
discrepancy is caused by an ambiguity in the convention of what is
called a localization length: The 'field theoretical' localization
length determines the exponential decay of the {\it average transition
probability} $\left \langle |G^+(\omega_+;l,l')|^2 \right
\rangle_{\omega_0} \sim \exp( - |l-l'|/\xi_o)$ between two remote
states $|l-l'|\gg\xi_o$.  In numerical measurements, however, one
computes the {\it average of individual decay constants}, which is to
say that one calculates $|l-l'|/\xi_n = \left \langle - {\rm ln}
|G^+(\omega_+;l,l')|^2 \right \rangle_{\omega_0}$. The lengths $\xi_o$
and $\xi_n$ thus defined do not coincide in general. According to
Ref.\cite{fm2} they are related by $\xi_o=4\xi_n$ for quasi
$1d$-wires. In view of this, our analytical result does agree with the
numerics.
To summarize, we have mapped both the unitary and the orthogonal
quantum kicked rotor on the supersymmetric nonlinear $\sigma$--model
for quasi $1d$ wires. This proves the longstanding conjecture that the
universal properties of these two classes of system are indeed the same.  
Our mapping is straightforward and direct and avoids some approximations
made in earlier work, namely (i) the replacement of the deterministic
rotor by a stochastic model and (ii) the passage from unitary to Hermitian 
randomness. Note that the rotor-metal analogy is not restricted to the 
phenomenon of strong localization.  
It has recently been shown that quantum interference
in metals manifests itself in various {\it pre--stages} of localization
such as non--trivial wavefunction statistics~\cite{fe,fm2} or the
appearence of pre--localized states\cite{km}. The exact correspondence
between quasi 1$d$ wires and the rotor suggests that these effects must be
observable in the latter, too. In fact, the rotor may be an ideal
model system for highly accurate numerical analyses of these
pre--localization phenomena, since it can be implemented more
efficiently on a computer than can weakly disordered multi--channel wires.
\par
We have benefitted from discussions with O. Agam, S. Fishman,
P. Freche, M. Janssen and A.Mirlin.

\end{multicols}

\begin{thebibliography}{99}
\bibitem{fn2} Note that this approximation is analogous to the
  expansion in powers of $\hbar/(p_F l)$ commonly employed in the
  study of disordered metals. Here $p_F$ is the Fermi momentum and $l$
  denotes the elastic mean free path, i.e. the minimum scale over
  which the fields of an effective field theory for these systems
  vary.
\end{thebibliography}
\end{document}